\documentclass[aps,twocolumn,prc,floatfix,showpacs,preprintnumbers,amsmath,amssymb,nofootinbib,groupedaddress]{revtex4-1}
\usepackage[output-decimal-marker={,}]{siunitx}
\usepackage{mathtools}
\usepackage[normalem]{ulem}
\usepackage{wasysym}
\usepackage{xcolor}
\usepackage{graphicx}
\usepackage{dcolumn}    
\usepackage{bigstrut}
\usepackage{amsmath,amsfonts,amsthm,bm}
\usepackage{CJK}
\usepackage[pdfstartview=FitH,
            CJKbookmarks=true,
            bookmarksnumbered=true,
            bookmarksopen=true,
            colorlinks,
            pdfborder=001,
            linkcolor=blue,
            anchorcolor=blue,
            citecolor=blue
            ]{hyperref}

\newcommand{\varR}{\vec r}

\newcommand{\coeffSch}{-\frac{\hbar^2}{2m}}

\def\be{\begin{equation}}
\def\ee{\end{equation}}
\def\beo{\begin{equation}}
\def\eeo{\end{equation}}
\def\bea{\begin{eqnarray}}
\def\eea{\end{eqnarray}}
\def\bse{\begin{subequations}}
\def\ese{\end{subequations}}

\begin{document}

\title{A first step in the nuclear inverse Kohn-Sham problem: from densities to potentials}

\author{G. Accorto}
\affiliation{Dipartimento di Fisica ``Aldo Pontremoli'', Universit\`a
degli Studi di Milano, Via Celoria 16, 20133 Milano, Italy}
\affiliation{Department of Physics, Faculty of Science, University of Zagreb, Zagreb, Croatia.}

\author{P. Brandolini}
\affiliation{Dipartimento di Fisica ``Aldo Pontremoli'', Universit\`a 
degli Studi di Milano, Via Celoria 16, 20133 Milano, Italy}

\author{F. Marino}
\affiliation{Dipartimento di Fisica ``Aldo Pontremoli'', Universit\`a 
degli Studi di Milano, Via Celoria 16, 20133 Milano, Italy}

\author{A. Porro}
\affiliation{Dipartimento di Fisica ``Aldo Pontremoli'', Universit\`a 
degli Studi di Milano, Via Celoria 16, 20133 Milano, Italy}

\author{A. Scalesi}
\affiliation{Dipartimento di Fisica ``Aldo Pontremoli'', Universit\`a 
degli Studi di Milano, Via Celoria 16, 20133 Milano, Italy}

\author{G. Col\`o}
\email{gianluca.colo@mi.infn.it}
\affiliation{Dipartimento di Fisica ``Aldo Pontremoli'', Universit\`a
degli Studi di Milano, Via Celoria 16, 20133 Milano, Italy}
\affiliation{INFN, Sezione di Milano, Via Celoria 16, 20133 Milano,
Italy}

\author{X. Roca-Maza}
\email{xavier.roca.maza@mi.infn.it}
\affiliation{Dipartimento di Fisica ``Aldo Pontremoli'', Universit\`a
degli Studi di Milano, Via Celoria 16, 20133 Milano, Italy}
\affiliation{INFN, Sezione di Milano, Via Celoria 16, 20133 Milano,
Italy}

\author{E. Vigezzi}
\email{enrico.vigezzi@mi.infn.it}
\affiliation{INFN, Sezione di Milano, Via Celoria 16, 20133 Milano,
Italy}

\date{\today}

\begin{abstract}
Nuclear Density Functional Theory (DFT) plays a prominent role in the understanding of nuclear structure, being the approach with the widest range of applications. Hohenberg and Kohn theorems warrant the existence of a nuclear Energy Density Functional (EDF), yet its form is unknown. Current efforts to build a nuclear EDF are hindered by the lack of a strategy for systematic improvement. In this context, alternative approaches should be pursued and, so far, an unexplored avenue is that related to the inverse DFT problem. DFT is based on the one-to-one correspondence between Kohn-Sham (KS) potentials and densities. The exact EDF produces the exact density, so that from the knowledge of experimental or {\it ab initio} densities one may deduce useful information through reverse engineering. The idea has already been proven to be useful in the case of electronic systems. The general problem should be dealt with in steps, and the objective of the present work is to focus on testing algorithms to extract the Kohn-Sham potential within the simplest ansatz from the knowledge of the experimental neutron and proton densities. We conclude that while robust algorithms exist, the experimental densities present some critical aspects. Finally, we provide some perspectives for future works.
\end{abstract}



\maketitle

\section{Introduction}

Density Functional Theory (DFT) has become gradually one of the best tools of choice for the study of nuclear structure \cite{BHF.03,book}, trying to follow the path that led to  the success of electronic DFT \cite{Burke.12,Becke.14}. There are analogies and differences between the two cases. One can expect that building an Energy Density Functional (EDF) for nuclei is harder than doing the same for electronic systems, in keeping with the more involved, and less well known, underlying nucleon-nucleon (NN) interaction. This interaction is strongly spin- and isospin-dependent, while momentum-dependent, spin-orbit and tensor terms are far from being negligible and there are also three-body (NNN) components --- all this is at variance with the Coulomb case. 

DFT is grounded in the Hohenberg-Kohn Theorems (HKTs), stating that a universal EDF must exist and yet not providing any guidance on how to build its terms \cite{HK.64}. The most used and well-established nuclear EDFs like the Skyrme and Gogny ones (we do not discuss covariant functionals which, though very successful, are outside our scope here) include terms that have their origin in central two-body forces and have a form proportional to the square of the total number density ($\rho^2$), repulsive terms which depend on a larger power of $\rho$ to mimic short-range repulsion, besides the terms that have been mentioned in the previous paragraph and account for spin forces, spin-orbit forces etc. They contain parameters that are fitted on experimental properties of selected nuclei, can be dubbed as phenomenological and lack from the beginning a clear mechanism for systematic improvement.

Recently, several groups have undertaken important steps to build more general EDFs, in which  one starts from $\rho$-dependent terms, and include other terms that depend on gradients $\vec \nabla \rho$ up to a given order (see \cite{PhysRevC.83.054311}, as well as \cite{PhysRevC.96.044330} and references therein). The systematic construction of all possible densities and their gradients  has been described in the past \cite{Engel:1975,DD:96}, together with  the systematic classification of all possible terms that should enter a nuclear EDF \cite{PhysRevC.69.014316}. These terms are all the scalar quantities that can be built out of densities and that are invariant under parity, time-reversal, rotational, translational and isospin transformations. As obvious, the number of such terms can be very large and fitting too general EDFs  may become either technically prohibitive, or unpractical due to the lack of experimental input, or both.

Other attempts have been made to derive the nuclear EDFs from  fundamental approaches. In this regard, general ideas and perspectives can be found in Ref.~\cite{Drut.10}. Whereas in Refs.~\cite{Cao.06,Baldo.08,Gam.11} attempts have been made to derive non-relativistic EDFs from Br\"uckner-Hartree-Fock calculations in uniform matter,  a hybrid approach has been followed in Refs.~\cite{PhysRevC.82.054307,PhysRevC.84.044306,PhysRevC.97.054304}, in which the long-range, pion exchange-like part of the EDF has been derived from chiral forces and short-range coupling constants are left to be fitted against phenomenological data. While trying to derive EDFs from {\it ab initio} theories, or to generalize their structure, may produce some breakthroughs, perhaps new mathematical or computational techniques are also worth attempting. 

The present work is based on the Kohn-Sham (KS) realization of the HKTs \cite{KS.65}. We define as direct problem the one in which, given a KS functional and the associated effective potential, we deduce the density that can be compared to experiment as well as the total ground-state energy. We define as inverse problem the one in which, starting from a given (supposedly exact) density, we deduce the effective KS potential. Solving the inverse problem is obviously appealing as it can constrain the phenomenological KS potential and the phenomenological EDF at its basis. In the case of electronic systems there have been several attempts to attack the inverse KS problem. In this work, we closely follow some of the inversion methods reviewed in Ref.~\cite{JW:18}, in which some basic concepts and techniques are discussed in detail. Hence, the current work is motivated by the idea that if the density is the basic variable to describe fermionic systems, as guaranteed by the HKTs, the nuclear densities should contain, in principle, all the relevant information to constrain the nuclear EDF.

The structure of the present paper is as follows. In Sec.~\ref{intro:iks}, 
we introduce the KS realization of the HKTs as well as the inverse problem, 
highlighting the specific aspects of the nuclear case as we have just mentioned. 
In Sec.~\ref{methods}, we describe the two adopted computational approaches. In Sec.~\ref{test}, we test the numerical methods presented in Sec.~\ref{methods} by using theoretical densities generated from a mean-field approach. In Sec.~\ref{result}, we deduce from experiment the KS potential for protons in ${}^{40}$Ca, and for neutrons and protons for ${}^{208}$Pb. 
Our conclusions will be presented in Sec.~\ref{conclusions}, together with some
future perspectives for this work.

\section{Inverse Kohn-Sham problem}
\label{intro:iks}

As is well known, the KS method is the most practical way to implement DFT. 
This method is based on the idea that solving the problem of the interacting system
under study is equivalent to solving a system of independent particles subject 
to an effective potential, provided the density is the same. In fact,
the ground-state number density
can be expressed as the sum over a number $N_{orb}$ of KS single-particle orbitals $\phi_i$, namely
\begin{equation}\label{eq:dens}
\rho(\vec r) = \sum_i^{N_{orb}} \vert \phi_i(\vec r) \vert^2.
\end{equation}
In addition, it is assumed that  the kinetic energy $T$ has the same form as in the 
independent particle case. 
Then, the EDF is written as \begin{equation}\label{KS_EDF}
E[\rho] = \sum_i \int d^3r\ \phi^*_i(\vec r) \left(
-\frac{\hbar^2}{2m} \right) \nabla^2\phi_i(\vec r) + F[\rho],
\end{equation}
and its minimization leads to the well-known Kohn-Sham equations of the type
\begin{equation}\label{eq:KS}
\left( -\frac{\hbar^2}{2m}\nabla^2 + \frac{\delta F[\rho]}{\delta\rho} 
\right) \phi_i(\vec r) = \varepsilon_i \phi_i(\vec r),
\end{equation}
where $\varepsilon_i$ are the so-called Kohn-Sham eigenvalues. The quantity  $\frac{\delta F[\rho]}{\delta\rho}$ plays the role of an effective potential and will be hereafter denoted by $U[\rho]$. If $F$ is assumed, $U$ is given and 
the density can be found. In this context, the IKS problem consists in reversing the procedure and deriving the effective potential $U[\rho]$ given the knowledge of $\rho$. As already mentioned, in the present work only   neutron and proton densities will be considered. That is, we will assume that the KS potential is a function of space coordinates 
through the dependence on the local densities 
only. This potential is determined except for a constant shift. In KS-DFT, eigenvalues and orbitals are auxiliary quantities which, strictly speaking, have no physical meaning. 
The exception is the energy of the 
last occupied orbital, that coincides with 
the ionization energy in atomic systems \cite{Perdew:1982, Perdew:1997}, 
or with the neutron/proton separation energy in nuclear systems. 
This provides a unique way to set the absolute value of the
KS potential, that  
goes to zero 
as it should.

In the literature, one finds several formulations of the IKS problem for the case of electronic systems, (see e.g.  \cite{JW:18} for a recent review and \cite{PhysRevA.47.R1591} for early references). The two approaches that we discuss below have been originally introduced in Refs.~\cite{PhysRevA.49.2421} and \cite{JW:18}. The IKS problem has been solved in the case of the He atom, where an exact analytic solution for wave functions and densities is available \cite{Hylleraas1929}, and this has allowed testing various approximate methods \cite{10.21468/SciPostPhys.6.4.040}. Among recent papers, we also would like to mention Refs.~\cite{Nielsen_2013,Nielsen2018}, that deal with the time-dependent inverse problem, and Ref.~\cite{Kumar}, that shows the connection between different IKS strategies. 
References \cite{Kanungo2019,Naito_2019} should also be highlighted. 
This list is not meant to be exhaustive. 
In nuclear physics, some groups have tried to deduce a heavy-ion potential from time-dependent Hartree-Fock
calculations \cite{PhysRevC.74.021601}.
Anyway, to the best of our knowledge, ours is the first attempt to address 
the IKS problem in the nuclear case. Therefore, we discuss some specific 
issues in what follows.

{\it Is the IKS problem well-posed?} One finds ample discussion in the 
literature regarding the question whether the IKS is well-posed, 
according to the definition given by  J. Hadamard \cite{Hadamard}, or not. 
As reported in \cite{JW:18}, according to J. Hadamard a problem is
well-posed if a solution exists, it is unique, and it depends
continuously on the data. If any of these three properties is violated then
the problem is ill-posed.
We miss a formal proof of the fact that IKS is well-posed, except for the case of discretized systems \cite{Chayes1985}, but this does not mean that the IKS problem is necessarily ill-posed. Still, its numerical solution is a very delicate matter. From a theoretical point of view, the existence of a Kohn-Sham potential  for a given experimental density is not guaranteed, although the uniqueness of the solution is guaranteed if a method converges (see, e.g., the discussion of not $v-$representable densities in \cite{Dreizler}). Furthermore, 
uncertainties and lack of completeness in the input density data can lead to violations 
of the Hadamard conditions that we have just mentioned. 
Some of the pathologies inherent to experimental nuclear densities will be discussed in Sec.~\ref{result}.

{\it EDF from KS potential.} Going one step back from the effective potential 
$U[\rho(\vec r)]$ to the functional $F[\rho(\vec r)]$ is also possible. 
In electronic systems (cf. e.g. Ref. \cite{Gaiduk}), it has been shown 
with concrete examples that the exchange-correlation functional 
$E_{\rm xc}$ can be reconstructed starting from the knowledge of
the associated exchange-correlation potential $v_{\rm xc}$. The basic
formula to achieve this goal has been introduced by R. van Leeuwen and 
E.J. Baerends in Ref. \cite{PhysRevA.51.170}, and reads
\begin{equation}
E_{\rm xc}[\rho] = \int_0^1 dt\ \int d^3r\ v_{\rm xc}[\rho_t] 
\frac{\partial\rho_t}{\partial t}.
\end{equation}
Here, $\rho_t$ is a continuously parametrized density such that 
$E_{\rm xc}[\rho_0]=0$ and $\rho_1=\rho(\vec r)$. This equation implies
the knowledge of the potential along a path of densities which, as 
we mentioned above, is not obtainable from any experimental input but
only from {\em ab initio} calculations. For example, calculations of 
systems of few nucleons confined in an external harmonic oscillator
potential (the so-called neutron drops) may provide a path of densities
as the confining potential can be varied continuously. 
We shall try to elaborate on this in the conclusions.

{\it Different types of nuclear densities.}
In the nuclear case, various kinds of densities must be used for a realistic EDF.
In fact, the classification of all possible
local densities that may (at least in principle) enter a local EDF, and the way to build them,
is discussed in detail in Refs. \cite{Engel:1975,DD:96,BHF.03}.
Consequently, our current inversion of Eq.~(\ref{eq:KS}) is intended as a first
step towards more realistic applications, as we employ here
neutron and proton number densities alone.
We shall need, in future, to formulate the inverse KS problem also
in terms of other
relevant nuclear densities. In this case, the implementation will
require the input from {\it ab initio} theories. We give just two
examples here. If we wish to include gradient terms in our considerations,
we could extract the density dependence from uniform matter and
formulate the IKS in finite nuclei to obtain the gradient terms only.
If we wish to include spin terms, we can formulate the IKS having
the spin density of odd nuclei from {\em ab initio} calculations.
We shall come back to this in the conclusions.

{\it Laboratory density and intrinsic density.} A fundamental difference between DFT in finite electronic systems and nuclei consists in the fact that in the former case the fixed ion positions constrain the shape of the system in the laboratory frame. In nuclei, that are self-bound systems, this is not the case. The usual HK theorem, as it has been argued by several authors \cite{Engel:2007}, is formulated for the laboratory density, while experiments on nuclei probe the intrinsic density (relative to the nuclear center of mass). Nonetheless, it has been proven that, given an arbitrary Hermitian operator $\hat Q$, one can build an energy functional depending on $Q(\vec r) \equiv \langle \hat Q(\vec r) \rangle$ that is universal in the HK sense and has its minimum at the correct value of $Q$ with the correct energy (see for example Sec. II of Ref.~\cite{Engel:2007} as well as Refs.~\cite{valiev1997,fernando2008}). In this respect, being the intrinsic density an Hermitian operator, one can replace the laboratory density with the intrinsic density in the HK theorem \cite{Barnea:2007,MBS:2009}.

\section{Inverse Kohn-Sham problem: two different methods}
\label{methods}

We start from Eq. (\ref{eq:KS}) and assume that the effective potential $U[\rho]$ is only position-dependent in keeping with the KS ansatz described in Sec.~\ref{intro:iks}. Thus, within this ansatz, non-local effects and the spin-orbit potential are not explicitly taken into account. The latter approximation should not impact much on the KS potential since spin-orbit effects are not expected to markedly change the KS orbitals. Consequently, from Eq.~(\ref{eq:dens}), the density should be almost untouched. Non-locality could be accounted via gradients of the density but, as we have said above, we leave this for future improvements. 
As for the Coulomb interaction between protons, we implicitly assume a local form within $U[\rho]$ in Eq.~(\ref{eq:KS}). This is known to work well for the description of the total binding energy 
and density in nuclei \cite{Titin:1974,AGDR2}.

We use two different methods to extract the KS potential $U[\rho]$ from the neutron and proton densities.
The first one is based on an iterative procedure and was introduced by R. Van Leeuwen and E.J. Baerends \cite{PhysRevA.49.2421}: it will be called vLB method. The second method consists instead in the constrained minimization of the kinetic energy $T$, in the spirit of KS that introduce an auxiliary system of independent particles with the same density of the system under study: this method will be called constrained variational, or CV, method. In choosing the notation vLB and CV, we follow Ref.~\cite{JW:18}. In both cases we will use $\tilde\rho$ to denote the target density, that is, the density to be reproduced. We will restrict ourselves to doubly-magic, spherical systems. 

\subsection{The vLB method}\label{subsec:vLB}

The aim of this procedure is to bring the calculated density as close as possible to the given target density by iteration. That is, by starting from an initial guess of $U[\rho]$,  one implements the vLB  algorithm \cite{PhysRevA.49.2421} to calculate the new  $U[\rho]$ and repeats  
it until $U[\rho]$ is stable. The derivation of the vLB algorithm is very simple so we outline it in what follows.

Let us start by writing the direct formulation of the KS equation (\ref{eq:KS}) in spherical symmetry, 
\begin{equation}
\biggl[ -\dfrac{\hbar^2}{2m}\dfrac{d^2}{dr^2} + \dfrac{\hbar^2l(l+1)}{2mr^2} + U(r) \biggr]u_i(r) = 
\varepsilon_i u_i(r),
\label{dir}
\end{equation}
where $u_i(r)$ are the reduced radial wave functions and $U(r)\equiv U[\rho(r)]$ is the effective Kohn-Sham potential that has been already introduced. 
 In the case of spherical nuclei, $i$ stands for {$n$, $l,j$} which  denote the principal quantum number,  the orbital angular momentum and the total angular momentum. The full wave function reads $\phi_i(\vec r) \equiv \frac{u_{nlj}(r)}{r} \left[ Y_l(\theta,\phi) \otimes \chi_{1/2} \right]_{jm}$, where $m$ denotes the projection of the angular momentum on the $z-$axis. We do not use specific notations for protons and neutrons as the iterative procedure is carried out independently for the two species. As already stressed, we assume here that the spin-orbit potential does not change significantly the shape of the radial wave function, so that $u_i(r)$ is effectively the same for the spin-orbit partners $j=l+1/2$ and $j=l-1/2$. 

The boundary conditions at the origin associated with Eq. 
(\ref{dir}) are well known,   
\begin{subequations}
\begin{equation}
\lim_{r\to0}u(r) = r^{l+1},
\end{equation}
\begin{equation}
\lim_{r\to0}u'(r) = (l+1)r^l,
\end{equation}
\label{bound}
\end{subequations}
and these allow to solve the {\it direct} Kohn-Sham equation (\ref{dir}) by means of a shooting algorithm. Thus, we find the energy eigenvalues and the eigenfunctions and derive the density $\rho(r)$ as
\begin{equation}
\rho(r) = \dfrac{1}{4\pi r^2}\sum_{i=0}^{N_{\rm orb}}n_i u_i^2(r),
\label{den}
\end{equation}
where $n_i$ is the occupation factor of the orbital $i$. In principle, $n_i$ could be taken as a fractional particle number (cf. Ref.~\cite{Perdew:1982}). However, we assume here $n_i = 2j+1$ since we limit ourselves to closed shell nuclei. We have checked that, not having implemented any spin-orbit effect in Eq.~(\ref{dir}), working within the uncoupled $l$ and $s$ or coupled $j=l+s$ schemes provides identical results.  

The vLB inverse algorithm introduced in Ref. \cite{PhysRevA.49.2421} 
(as well as a slightly different algorithm~\cite{PhysRevA.47.R1591} that is not discussed here) can be obtained by algebraic manipulation of Eq. (\ref{dir}). Specifically, by 
multiplying Eq. (\ref{dir}) by $n_iu_i(r)$ at both sides, summing over $i$ and dividing by $4\pi r^2\rho(r)$, one finds
\begin{eqnarray}\label{eq7} 
U(r) & = & \frac{1}{4\pi r^2\rho(r)} \sum_{i=0}^{N_{\rm orb}} 
\left[ n_i u_i(r) \left( \frac{\hbar^2}{2m}\frac{d^2}{dr^2} -U_l \right) u_i \right. \nonumber \\
& + & \left. \varepsilon_i n_i u_i^2 
\right],
\end{eqnarray} 
where $U_l$ is a shorthand notation for the centrifugal potential. Now we need to define the iterative process from that equation. Denoting iteration numbers by superscripts, a guess for the new potential $U^{(k+1)}$ can be obtained substituting 
$\rho$ with $\tilde{\rho}(r)$ in the denominator at the r.h.s. of Eq.~(\ref{eq7}), 
and realizing that the rest of the r.h.s in the same equation corresponds to the potential times the density determined in the previous iteration, or $4\pi r^2\rho^{(k)}(r) U^{(k)}$. That is, 
\begin{eqnarray}
U^{(k+1)}(r) & = & 
\frac{1}{4\pi r^2\tilde\rho(r)} \sum_{i=0}^{N_{\rm orb}} 
\left[ n_i u_i^{(k)}(r) \left( \frac{\hbar^2}{2m}\frac{d^2}{dr^2} -U_l \right) u_i^{(k)} \right. \nonumber \\
& + & \left. \varepsilon_i n_i \left( u_i^{(k)} \right)^2 
\right] = \frac{\rho^{(k)}(r)}{\tilde\rho(r)}U^{(k)}(r).
\label{inv_nuc_simple}
\end{eqnarray}
It is important to note that Eq.~(\ref{inv_nuc_simple}) has a simple meaning. 
 In regions where the density at the $k$-th step is larger than the target density the potential is increased in absolute value, and vice versa. This makes sense for repulsive potentials as in the electronic case, but in the case of the attractive potentials needed in nuclear physics, the opposite should happen. %
To avoid this problem, we have adopted a modified algorithm proposed in Ref.~\cite{JW:18}:
\begin{equation}\label{inv_best}
U^{k+1}(r) = U^{(k)}(r) +\gamma\frac{\rho^{(k)}(r)-\tilde{\rho}(r)}{\tilde{\rho}(r)} \ .  
\end{equation}
This algorithm works equally well for attractive and repulsive potentials and  is invariant under an arbitrary shift of the potential,
providing a convenient  numerical alternative. Furthermore, we  have found that the 
simple choice $\gamma= 1$ MeV leads to consistent
results.

In short, our choice has been that of assuming 
a starting potential $U^{(k=1)}(r)$,    calculating the eigenstates  $u^{(k=1)}(r)$ 
from Eq.~(\ref{dir})
together with  the density (\ref{den}), then finding  a new potential 
by applying Eq.~(\ref{inv_best}). The procedure is repeated until 
convergence. 
The convergence condition used to stop the iterative procedure is set in terms of the absolute variation of the potential, that is, 
\begin{equation}
\Delta U^{(k)} \equiv \max_r \biggl|U^{(k+1)}(r) - U^{(k)}(r)\biggr| < \alpha \ .
\label{eq:alpha}
\end{equation}
Remarkably, despite its simplicity, the algorithm has proven to be robust enough to converge to the same results 
assuming $U^{(k=1)}(r)$ to be equal 
either to a realistic Woods-Saxon potential or to a 
simple
constant. 

\subsection{The CV method}\label{subsec:CV}

In the CV method, the IKS problem is formulated as a variational problem. The formulation is in keeping with the KS ansatz, that asserts that for any interacting Fermi system one can always postulate an independent particle system with the same density. Accordingly, in the CV method one writes down the kinetic energy functional of the $N$ fermions with the purpose of minimizing it. Therefore, in what follows, we shall use the name of objective functional for the expectation value of the kinetic energy associated with the wave function of $N$ independent particles as in Eq. (\ref{KS_EDF}). The minimization is subject to the following constraints:
\begin{enumerate}
\item the single particle orbitals $\phi_i(\vec r)$ must be orthonormal;
\item the density of the system $\rho(\vec r)$ must 
be equal to the target density $\tilde\rho(\vec r)$, for each value of $\vec r$. 
\end{enumerate}

Let us start from the same assumptions as in the previous subsection. We assume that 
we have $N_{\rm orb}$ single particle states and that  each of them has occupancy $n_i$. We prefer here to write the equation without going immediately to the spherically symmetric case, as the CV method is more apt for the generalization to the case in which this symmetry is totally or partially broken. The objective functional reads
\begin{equation}\label{eq: def funzione obiettivo con laplaciano}
T _s[\{\phi_{i}\}] = \sum_{i=1}^{N_{\rm orb}} n_{i} \int d^3r\ \phi_i^*(\varR) \left( \coeffSch  \right) \nabla^2{\phi_i(\varR)} \ .
\end{equation}
The orthonormality of the orbitals is the first constraint and is expressed by
\begin{equation}\label{eq: vincolo normalita integrale}
G_{ij} \left[ \left\lbrace  \phi_i \right\rbrace    \right] \,  =
\int d^3r\ \phi_i^*(\varR) \phi_j(\varR) = \delta_{ij}.
\end{equation}
The equality of the density to the target density represents the second constraint 
[see Eq.~(\ref{eq: def funzionale costo base})]. 
We introduce Lagrange multipliers that correspond to these constraints. Those associated with the constraints (\ref{eq: vincolo normalita integrale}) are indicated as $\varepsilon_{ij}$,  while we write the constraint associated with the density as $U(\varR)$. The constrained minimization of the objective functional is re-written as the free minimization of the functional that includes the Lagrange multipliers, which is named cost functional \cite{wu2003}. This cost functional reads
\begin{widetext}
\begin{equation}
\label{eq: def funzionale costo base}
	J\left[ \left\lbrace \phi_i \right\rbrace; U(\varR), \, \left\lbrace \varepsilon_{ij} \right\rbrace   \right] \, = \, 
	T_s \left[ \left\lbrace \phi_i \right\rbrace \right] \, +
	\int d^3r\ U(\varR) \rho(\varR)  \, - \, \sum_{i=1}^{N_{\rm orb}} \sum_{j=1}^{i} 
	\varepsilon_{ij} \int d^3r\ \phi_i^{*}(\varR) \phi_j(\varR) \ ,
\end{equation}
\end{widetext}
and the CV method consists in solving the equation 
\begin{equation}\label{vareq}
  \delta J\left[ \left\lbrace \phi_i \right\rbrace; U(\varR), \, \left\lbrace \varepsilon_{ij} \right\rbrace   \right] = 0 \ .
\end{equation}  

It is important to note that, as explained at p. 146 of Ref. \cite{parr1994density}, 
the optimization of the cost functional (\ref{vareq}) leads to a non-diagonal form of the 
KS equations. In other terms, the wave functions $\phi_i$ that are solution of (\ref{vareq}) are 
not eigenfunctions of the KS equations and $\varepsilon_{ii}$ do not correspond to the KS 
eigenvalues. Nevertheless, the set $\phi_i$ and the matrix
$\varepsilon_{ij}$ are related to the orbitals that solve the KS equation and 
to the diagonal matrix with the KS eigenvalues, respectively, by a unitary transformation. 
At the same time, the very form of Eq. (\ref{eq: def funzionale costo base}) tells that the 
extracted Lagrange multiplier $U(\vec r)$ is the Kohn-Sham potential, up to a constant shift. 
In this respect, the CV method is a direct formulation of the IKS.

Minimizing the cost functional $J\left[ \left\lbrace \phi_i \right\rbrace; U(\varR), \, \left\lbrace \varepsilon_{ij} \right\rbrace   \right]$ means to find the set of orbitals $\phi_i$, defined on a given domain, that gives the minimum value for the functional respecting the two constraints at the same time. Hence, an unrestricted three-dimensional formulation is technically involved. In order to compare to our results obtained with the vLB method presented in Sec.~\ref{subsec:vLB}, we have limited ourselves to the case of spherical nuclei, as already discussed, so that the problem becomes one-dimensional, 
and  the orbitals depend only on $r$. In spherical symmetry, Eq.~(\ref{eq: def funzionale costo base}) becomes 
\begin{widetext}
\begin{equation}
	J\left[ \left\lbrace u_j \right\rbrace; U(r), \, \left\lbrace \epsilon_{ij} \right\rbrace   \right] \, = \, T_s \left[ \left\lbrace u_j \right\rbrace \right] \, +
	4 \pi  \int_{0}^{\infty} U(r) \rho(r) r^2 \, dr \, - \, \sum_{i=1}^{N_{orb}} \sum_{j=1}^{i} \epsilon_{ij} \delta_{l_i l_j}  \delta_{j_i j_j}
	\int_{0}^{\infty} u_i(r) u_j(r) \, dr
\end{equation}
\end{widetext}

The CV method has been implemented following some important modifications suggested in Ref.~\cite{JW:18}. First, a new set of variables, {\it viz.} the re-scaled orbitals $f_i(r)$ that are defined by 
\begin{equation}\label{eq_def_f_riscalate}
u_i(r) = \sqrt{4\pi \tilde\rho(r) } \, \, r f_i(r) \ ,
\end{equation}
have been introduced. 
The rationale behind this substitution is that we expect that $u_i^2(r) \approx 4\pi r^2 \tilde\rho(r)$, so that the re-scaling will produce new functions $f_i(r)$ of the order of $\approx 1$, characterized by a milder behavior as compared  to $u_i(r)$. This helps in reducing the rounding errors that appear when operating  with quantities of different orders of magnitude. The CV iterative procedure starts with a guess for the wave functions, 
and not for the KS potential, at variance with 
the vLB procedure. Also in the case of the CV method, we have checked that the starting guess is not influencing the result of the minimization. Indeed, starting either 
from constant wave functions or from harmonic oscillator orbitals (with the usual 
$\hbar\omega=41\ {\rm A}^{-1/3}$ MeV), we obtain satisfactory results. 

The optimization of Eq.~(\ref{vareq}) is performed employing the IPOPT library \cite{Wachter2006,intro_IPOPT}. Two main conditions determine the convergence of the algorithm:
\begin{enumerate}
\item the relative tolerance on the violation of the constraints. This means that, at each step during the optimization, there is a test of the condition
\begin{equation}
\max_i \, | \frac{g_i-c_i}{g_i} |< \epsilon,
\label{eq:epsilon}
\end{equation}
where the constraints are  denoted by $c_i$,  and the quantities subject to the constraints are  denoted by $g_i$.
\item the tolerance on the value of the objective function $f$. The algorithm stops if the change in value of the objective function between two successive iterations $k$ and $k-1$  is smaller than a given tolerance $\delta$, namely
\begin{equation}
|f^{(k)}-f^{(k-1)}|\, <\, \delta,
\label{eq:delta}
\end{equation}
\end{enumerate}

Notice that while 
the CV method checks the convergence of the objective function $f$ [Eq.~(\ref{eq:delta})], that is, of the re-scaled radial wave functions $u_i(r)$, the convergence criteria of the vLB method [Eq. (\ref{eq:alpha})] is based on the change of the KS potential $U[\rho]$.

The transformation in Eq. (\ref{eq_def_f_riscalate}), that ensures numerical stability 
and accuracy in the solution of Eq. (\ref{vareq}), 
has the effect that the Lagrange multipliers obtained with IPOPT have lost their direct 
and clear connection with $U(\vec r)$ and $\varepsilon_{ij}$. 
Therefore, in a second step, we have written explicitly 
the differential equations that correspond to the variation of the cost functional. Since at this
stage the auxiliary wave functions at the constrained minumum are known, the equations become a set of
algebraic equations that is easily solved to obtain
$U(\vec r)$ and $\varepsilon_{ij}$ (cf. Ref. \cite{JW:18}). 

\section{Test of the numerical methods}
\label{test}

\begin{figure*}[t!]
\includegraphics[width=0.48\linewidth,clip=true]{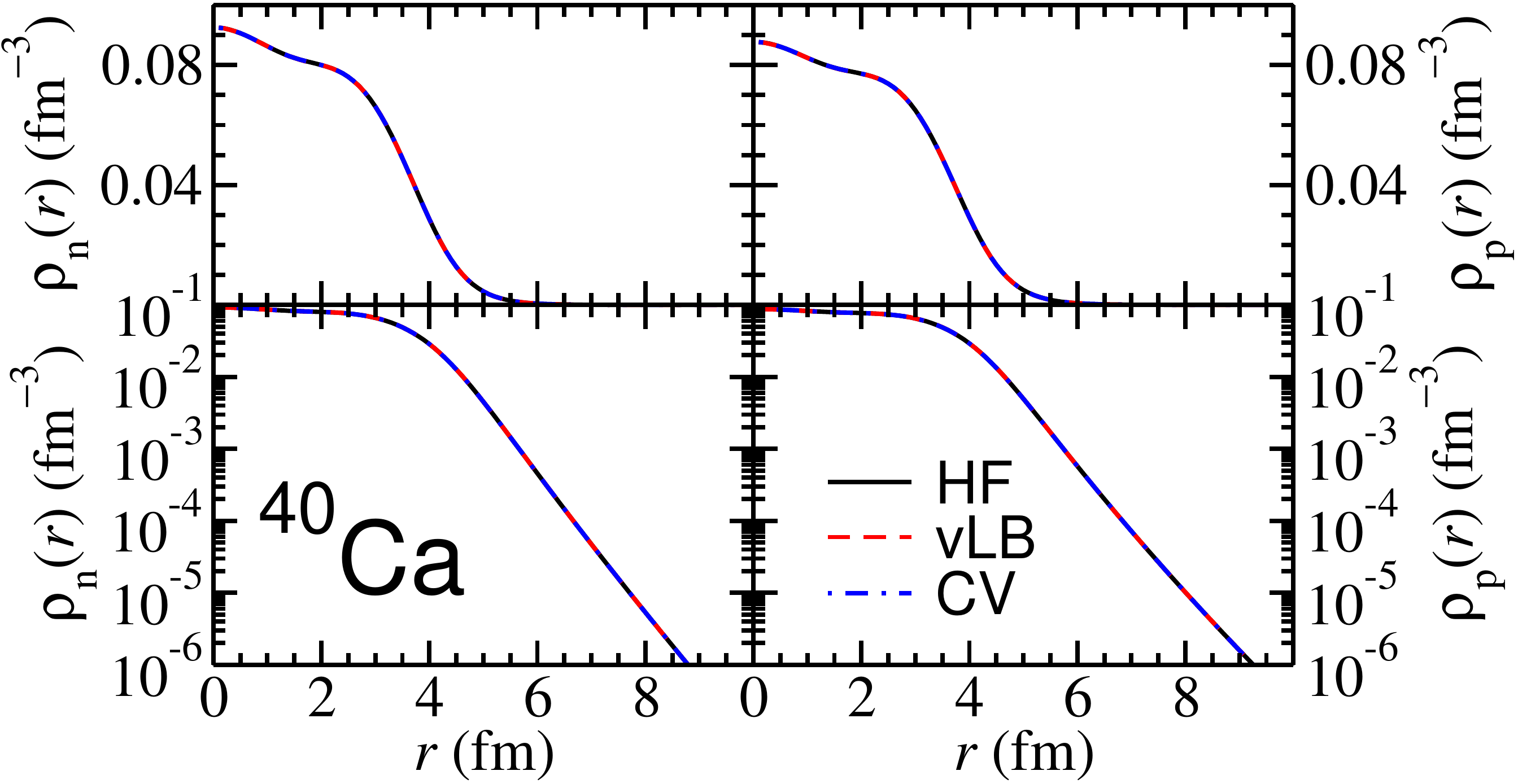}
\hfill
\includegraphics[width=0.48\linewidth,clip=true]{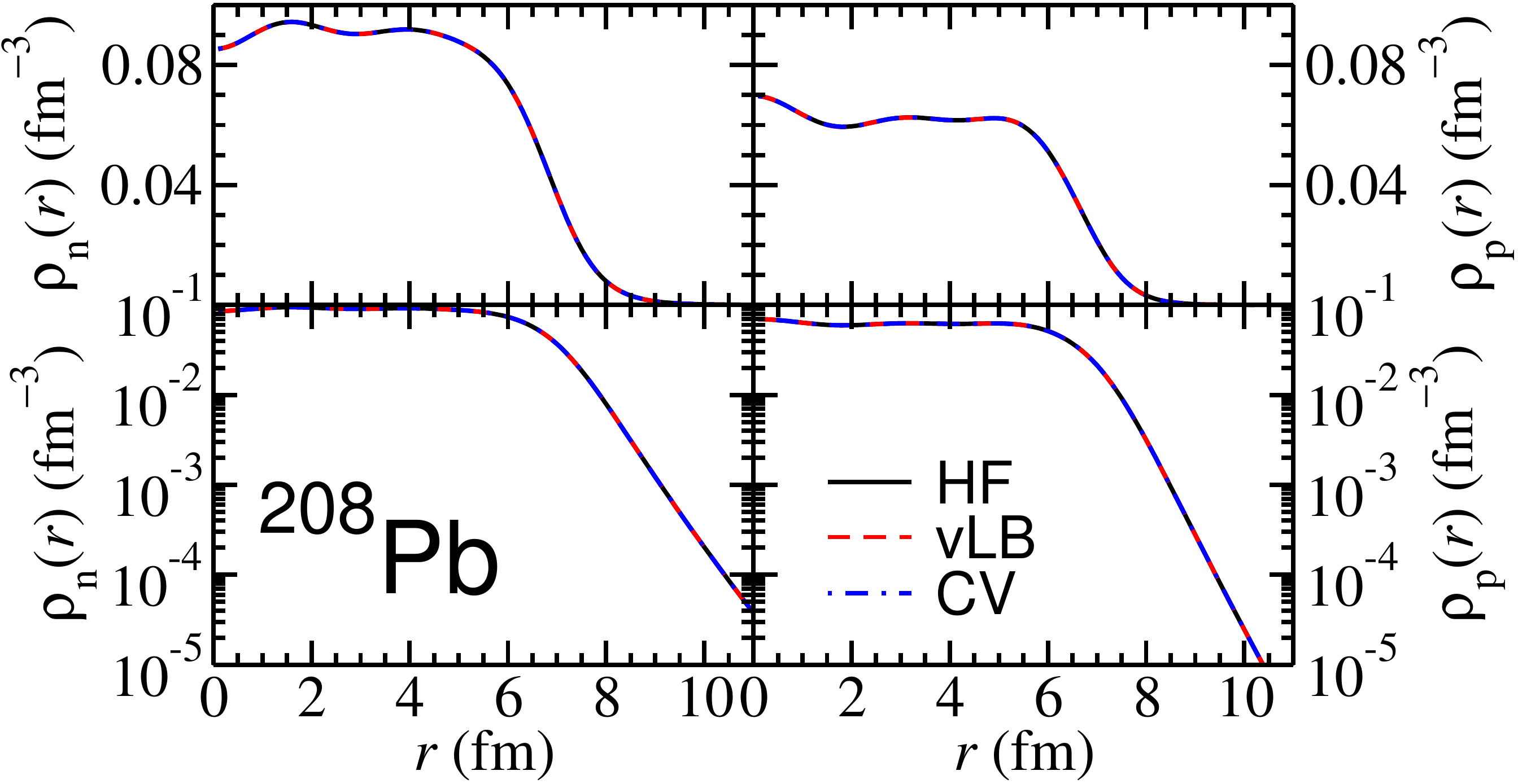}
\caption{(left figure) The neutron density, $\rho_n(r)$, and  the proton density, $\rho_p(r)$, in $^{40}$Ca 
are displayed as a function of the radial coordinate, in linear scale
(top panels) and in logarithmic scale (bottom panels).
The target densities obtained from HF calculations based on the SkX functional \cite{SkX}  (black solid line) are compared  with the 
densities resulting  from the two inversion methods discussed in the main text, namely  vLB (red dashed line) and CV (blue dot-dashed line).
(right figure) The same for the nucleus $^{208}$Pb.}
\label{fig:Den_HF}
\end{figure*}

\begin{figure*}[t!]
\includegraphics[width=0.48\linewidth,clip=true]{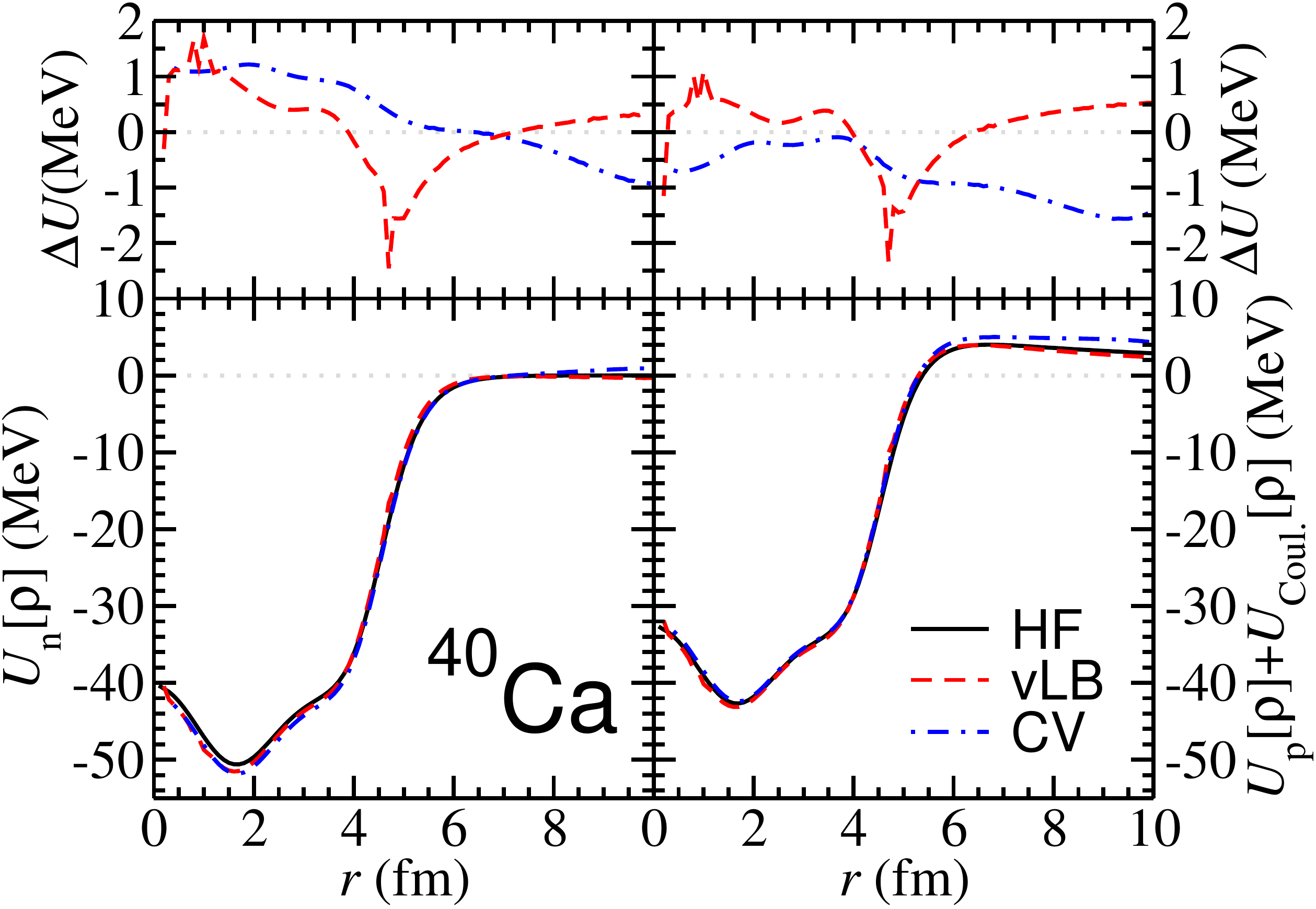}
\hfill
\includegraphics[width=0.48\linewidth,clip=true]{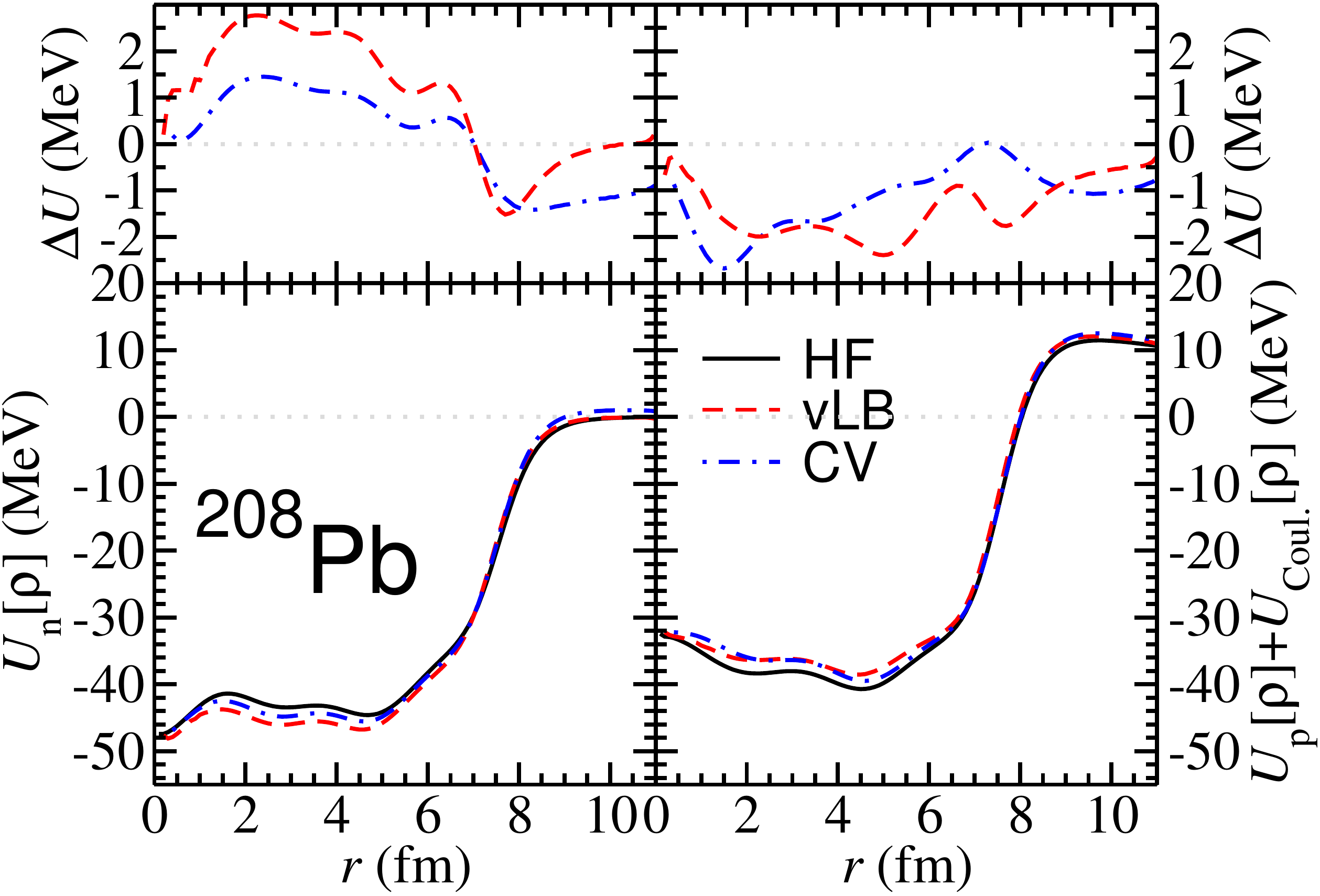}
\caption{(left figure) The Kohn-Sham potentials  for neutrons and protons obtained with the vLB (red dashed lines) and CV (blue dot-dashed lines) inversion methods in $^{40}$Ca are displayed  in the bottom panels as a function of the radial coordinate.  The benchmark HF calculations based on the SkX functional \cite{SkX} are also shown (black solid lines). 
In the top panels, the differences $\Delta U\equiv U_{\rm method}-U_{\rm HF}$ between  the vLB and CV potential 
and the  HF potential are shown. (right figure) The same, for the nucleus $^{208}$Pb.}
\label{fig:Pot_HF}
\end{figure*}

In this section, we test the two methods described in Sec.~\ref{methods} by using target densities produced by Hartree-Fock (HF) calculations in the doubly-magic nuclei $^{40}$Ca and  $^{208}$Pb. The HF calculations have been carried out by using the Skyrme interaction SkX \cite{SkX}. 
In the case of  the Skyrme interactions the HF equations differ from  the 
direct KS equations  for the presence of the spin-orbit potential 
and  of the effective mass $m^*(r)$.
The non-localities introduced through effective masses are very mild  in the case of the SkX interaction.   
For instance, the ratio $m^*/m$  in $^{208}$Pb lies  between  0.92 and 1  and between 1 and 1.08, for  protons and neutrons respectively.
We then expect that the inverse  KS algorithms  lead to potentials  which are similar to the SkX-HF potentials.

In Fig.~\ref{fig:Den_HF}, the neutron (left) and the proton (right) target densities from the HF calculations (black solid lines) in $^{40}$Ca (left) and $^{208}$Pb (right) are shown. Upper panels show the different densities in a linear scale while the lower panels show the same quantities in a logarithmic scale. The logarithmic scale is important to check the asymptotics of the densities. The results of the inversion methods vLB (red dashed lines) and CV (blue dash-dotted lines) reproduce in very much detail the target HF densities. The maximum and average values (with respect to the radial coordinate) of the absolute differences are shown in Table \ref{table_diff_SkX}. In short, the reproduction of the target densities is fully satisfactory, for both neutrons and protons, with either method, in these two doubly-magic nuclei. Nevertheless, the convergence criteria in the CV method are more restrictive in these calculations, as it is evident from the same table.  
\begin{table}[h!]
\centering
\begin{tabular}{r|cc|cc}
\hline\hline
Nucleus   & \multicolumn{2}{c|}{vLB}&\multicolumn{2}{c}{CV}\\
          &Max. & Aver. &Max. & Aver. \\
\hline
$^{40}$Ca~(p)  & 8.0 & 1.3&  1.0 & 0.2\\
$^{40}$Ca~(n)  & 8.7 & 1.4&  1.0 & 0.2\\
$^{208}$Pb~(p) & 2.6 & 0.3&  0.4 & 0.1\\
$^{208}$Pb~(n) & 8.7 & 3.6&  7.5  & 2.1 \\
\hline\hline
\end{tabular}
\caption{Maximum (Max.) and average (Aver.) differences between the target HF (SkX) and KS neutron (n) and proton (p) densities from the two inversion methods for the case of $^{40}$Ca and $^{208}$Pb. Numerical values are all in $10^{-6}~\si{fm^{-3}}$.}
\label{table_diff_SkX}
\end{table}

The Kohn-Sham potentials obtained with the two inversion methods are shown 
in Fig.~\ref{fig:Pot_HF}. As mentioned, the potentials are obtained 
up to an arbitrary constant. 
As explained in Sec. II, it is enough to shift the potential obtained 
through IKS so that the last 
occupied KS eigenvalue coincides with the last occupied HF eigenvalue to obtain such a constant. 
 
Despite numerical errors, the results for the effective potentials appear to be rather satisfactory.
The absolute value of the resulting difference $\Delta U$ with respect to the HF
potentials is smaller than 2.5 MeV both for protons and neutrons
(cf. the upper panels in Fig.~\ref{fig:Pot_HF}). 
This appears to be a quite reasonable agreement. 
We note that again the CV method seems to perform slightly better than the vLB method.
The reason stems from the different convergence criteria. The spin-orbit energy splittings, 
which exist in HF and are not taken care of in our procedure, have been 
checked to have no special influence (it is  
well known that spin-orbit shifts do not affect wave functions and densities, as a rule).
We recall here that there is another source of difference between the HF eigenvalues 
and the KS eigenvalues. While the former contain some effects due to the effective mass 
$m^*/m \lesssim 1$, the latter assumes $m^*/m = 1$. 
Thus, we note a small deviation of the KS potentials in their asymptotic behavior 
\cite{Perdew:1997} for $r\rightarrow \infty$. 

\begin{figure}[hbt]
\includegraphics[width=\linewidth,clip=true]{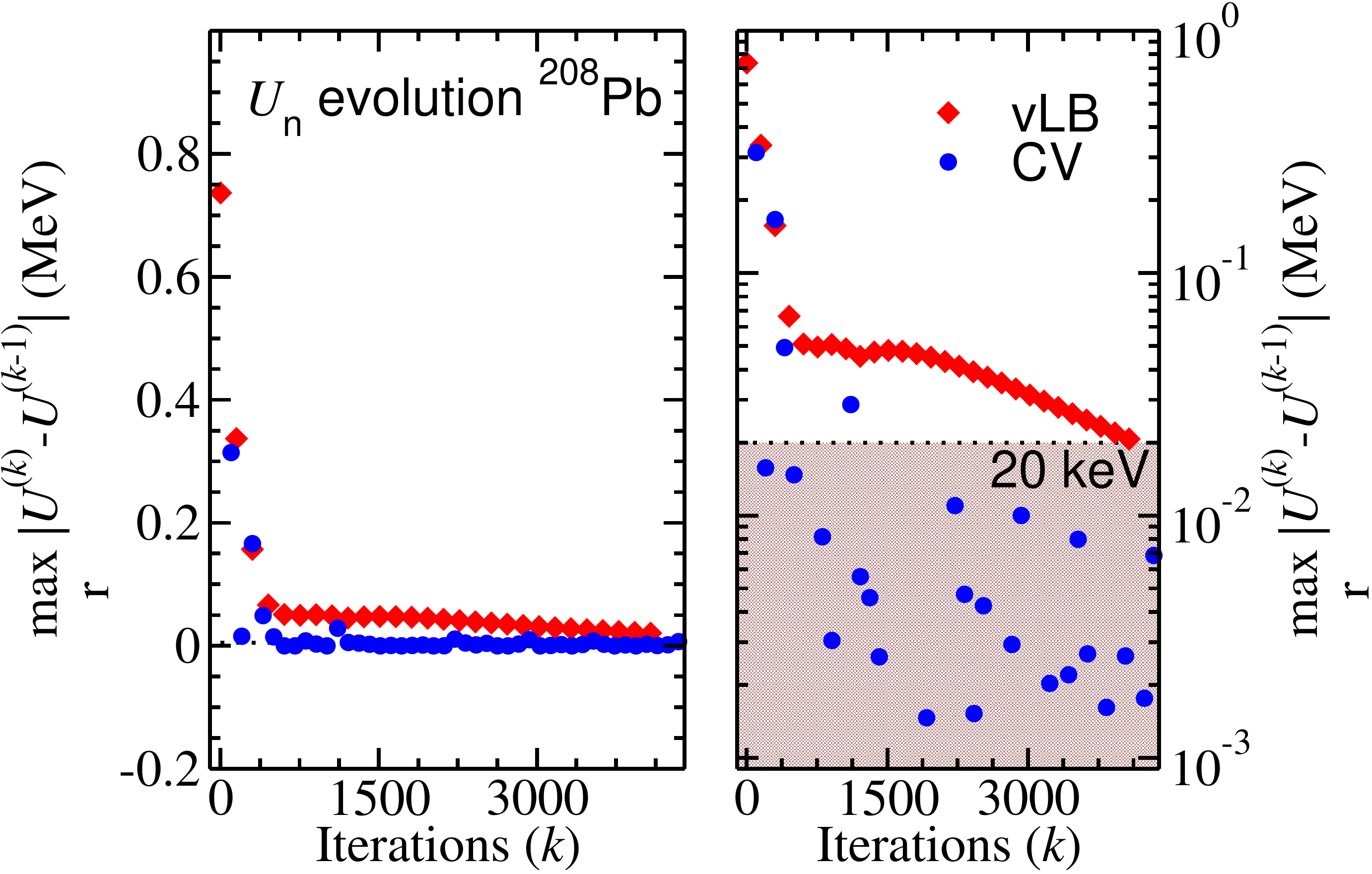}
\caption{Convergence test of the inversion methods, using as target density the results of a HF calculation  with the SkX interaction. 
The absolute difference $\Delta U^{(k)}$ [cf. Eq. (\ref{eq:alpha})] between the neutron Kohn-Sham potentials for $^{208}$Pb 
calculated at two successive iterations  is shown at different iteration steps on a linear (left panel) and  on a logarithmic (right panel) scale.  
Results of the inversion method vLB and CV are shown by red diamonds and blue circles respectively. 
The value of 20 keV associated with the convergence criterion of the vLB method, discussed
in the main text, is highlighted.
} 
\label{fig:Convergence}
\end{figure}

We now focus on the convergence of the procedures. The two algorithms behave in a quite different way. As explained in Sec.~\ref{subsec:vLB}, the vLB method iterates the potential according to Eq.~(\ref{inv_best}) and stops when the condition 
(\ref{eq:alpha}) is satisfied: in the present case, we set $\alpha =20$ keV, and  the iteration procedure is stopped when 
$\Delta U^{(k)} \le\SI{20}{keV}$. In Fig.~\ref{fig:Convergence}, we display the evolution of the quantity $\Delta U^{(k)}$ 
as a function of the number of iterations for the case of the neutrons in $^{208}$Pb, on a linear scale (left panel)  and on a logarithmic scale (right panel)
For the sake of clarity, only a representative point every 150 iterations is shown. 
The values of  $\Delta U^{(k)} $  obtained  with the vLB method (shown by red diamonds) decrease rapidly from 1 to about 10$^{-1}$ during the first 500 iterations; subsequently, the decrease continues but at a much slower pace.
The results from the CV method (shown by blue dots)  should be seen under a different light. The procedure does not use the 
quantity $\Delta U^{(k)}$ as a convergence criterion, but 
 attempts to minimize the value of the kinetic energy taking into account the tolerance with which constraints should be fulfilled. Then, the 
 values of  $\Delta U^{(k)}$ corresponding to different iterations do not, and should not be expected to, decrease with a monotonic trend. 
 They show instead  an oscillatory behavior, which can be observed looking at the right panel in logarithmic scale.  With this caveat 
 we nevertheless observe that the quantity $\Delta U^{(k)}$ shows an overall  decreasing trend as the iteration process goes on, 
 becoming small enough to conclude that the final result for the potential is indeed reliable.

\section{Results for experimental densities}
\label{result}

In this section, we extract KS potentials from the experimental densities. As case studies, we use the proton density of $^{40}$Ca and the proton and neutron densities of $^{208}$Pb. Specifically, the proton densities come from the electron scattering data of Ref.~\cite{DeVries}, while the neutron density of $^{208}$Pb has been 
extracted from proton scattering measurements in Ref.~\cite{Zenihiro}. In both references, a parameterization of the electromagnetic charge and neutron densities based on a sum of Gaussian functions (SoG) can be found. This method was first introduced in Ref.~\cite{sick1974} to extract nuclear charge densities from elastic electron scattering data without using model distributions but a basis of well behaved functions. 

The charge density distribution expressed as SoG can be written as follows, 
\begin{equation}\label{eq:SOG}
\rho_{\rm charge}(r) \, = \sum_i A_i^{(\rm charge)} \left( e^{- \left( \frac{r-R_i}{\gamma}\right)^2 }  +   e^{- \left( \frac{r+R_i}{\gamma}\right)^2 }    \right).
\end{equation}
The coefficients $A_i^{\rm charge}$ are given by
\begin{equation}
\label{eq: coeff SOG prot}
A_i^{\rm charge} \, = \, \frac{Ze Q_i}{2 \pi^{\frac{3}{2}} \gamma^3   \left( 1 + \frac{2R_i^2}{\gamma^2}\right) },
\end{equation}
where $Q_i$ is the fraction of total charge that is associated with the integral of the $i$-th Gaussian. Accordingly, the normalization condition $\sum_i Q_i = 1$ holds. The Gaussians are centered at different points $R_i$, whereas the widths are characterized by a common value $\gamma$. Sometimes, the value of $\gamma$ is taken  to be close to the width of the narrowest peak that one finds when inspecting the square of typical Hartree-Fock or Woods-Saxon wave functions for the nucleus under study. The reason why the SoG parameterization (\ref{eq:SOG}) is chosen is that, if the sum contains enough terms, it corresponds to a model-independent representation of the actual data points. In principle, this would require a very large number of Gaussian terms if the experimental data could cover the full momentum transfer range, that is from 0 to infinity. In practice this is not the case and, thus, a manageable number of terms, of the order of 10-15, has been proven to be stable against small changes. As it can be easily understood, this representation may suffer from the fact that experimental data is scarce  or does not cover a wide enough range of beam energies and scattering angles. 

In order to extract the proton densities from the charge densities, we have neglected the small effects due to the electromagnetic spin-orbit and the neutron electromagnetic finite size (see for example Sec. II.B of Ref.~\cite{ray1979}). Hence, we have extracted the proton densities from the charge densities as follows, 
\begin{equation}
\rho_{\rm charge}(\vec r) = \int d^3r'\ f(\vec r^\prime)\rho_p(\vec r-\vec r^\prime) \ , 
\end{equation}
using the approximate electric proton form factor 
\begin{equation}
f(\vec r) \, = \frac{q_e}{\pi^{3/2}\alpha^3} e^{-\left( \frac{r}{\alpha} \right)^2 },
\end{equation}
where 
$q_e$ is the proton charge and
$\alpha$ has been taken as $\sqrt{2/3} R_p$ and the value $R_p=0.87$ fm has been assumed for  the r.m.s. proton radius. Small changes on the chosen value for $R_p$ will not appreciably change our results. The deconvolution that leads to the proton charge density is performed using the regular product in Fourier space. Due to the properties of the Gaussian functions, the result in coordinate space can be analytically written assuming spherical symmetry as 
\begin{widetext}
\begin{equation}
\label{eq: densita SOG protone}
\rho_p(r) \, = \, \sum_i  \frac{ \gamma^3 A_i}{e \beta r} \left[     \left( \frac{r-R_i}{\beta^2} + \frac{R_i}{\gamma^2} \right) e^{-\left(  \frac{r-R_i}{\beta}  \right)^2 }  +
\left( \frac{r+R_i}{\beta^2} - \frac{R_i}{\gamma^2} \right) e^{-\left(  \frac{r+R_i}{\beta}  \right)^2 }
\right],
\end{equation}
\end{widetext}
where $\beta = \sqrt{\gamma^2 - \alpha^2}$.

In the case of neutrons, such procedure is not needed as Ref.~\cite{Zenihiro} provides the neutron density in the form of Eq. (\ref{eq:SOG}) directly. These data have been obtained via proton elastic scattering. Protons interact via the strong interaction with both neutrons and protons. So if the proton density is known, one can derive the neutron density compatible with the experimental cross section. This procedure is not model-independent, 
at variance with the case of electron elastic scattering used 
to determine the electromagnetic charge density. In fact, the proton-nucleus interaction
at intermediate incident energy is relatively well-known but has some uncertainty.  

\begin{figure}[b!]  
\includegraphics[width=\linewidth,clip=true]{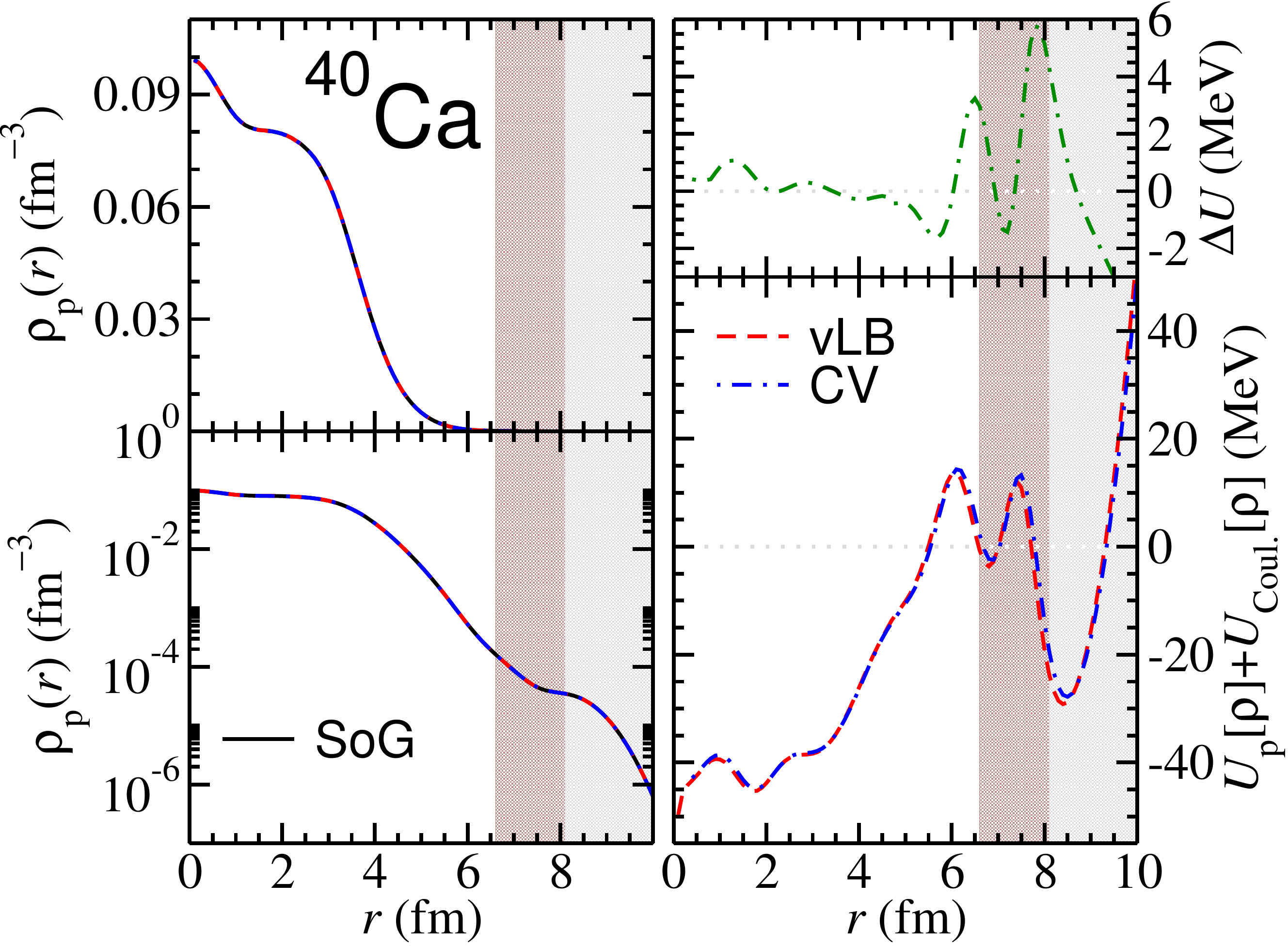}
\caption{The proton density for the case of ${}^{40}$Ca is displayed as a function of the radial coordinate on a linear scale (top left panel) and on a logarithmic scale (bottom  left panel). The target experimental density \cite{DeVries} labeled  SoG -- sum of Gaussians -- (black solid lines) is compared with those obtained with the inversion methods vLB (red dashed lines) and CV (blue dot-dashed lines). In the bottom right panel, the Kohn-Sham potentials obtained within the two inversion methods are compared, and in the  top right panel their difference $\Delta U=U_{\rm CV}-U_{\rm vLB}$ is shown.}
\label{fig:40Ca_Den_Pot}
\end{figure}

\begin{figure*}[hbt]
\includegraphics[width=0.48\linewidth,clip=true]{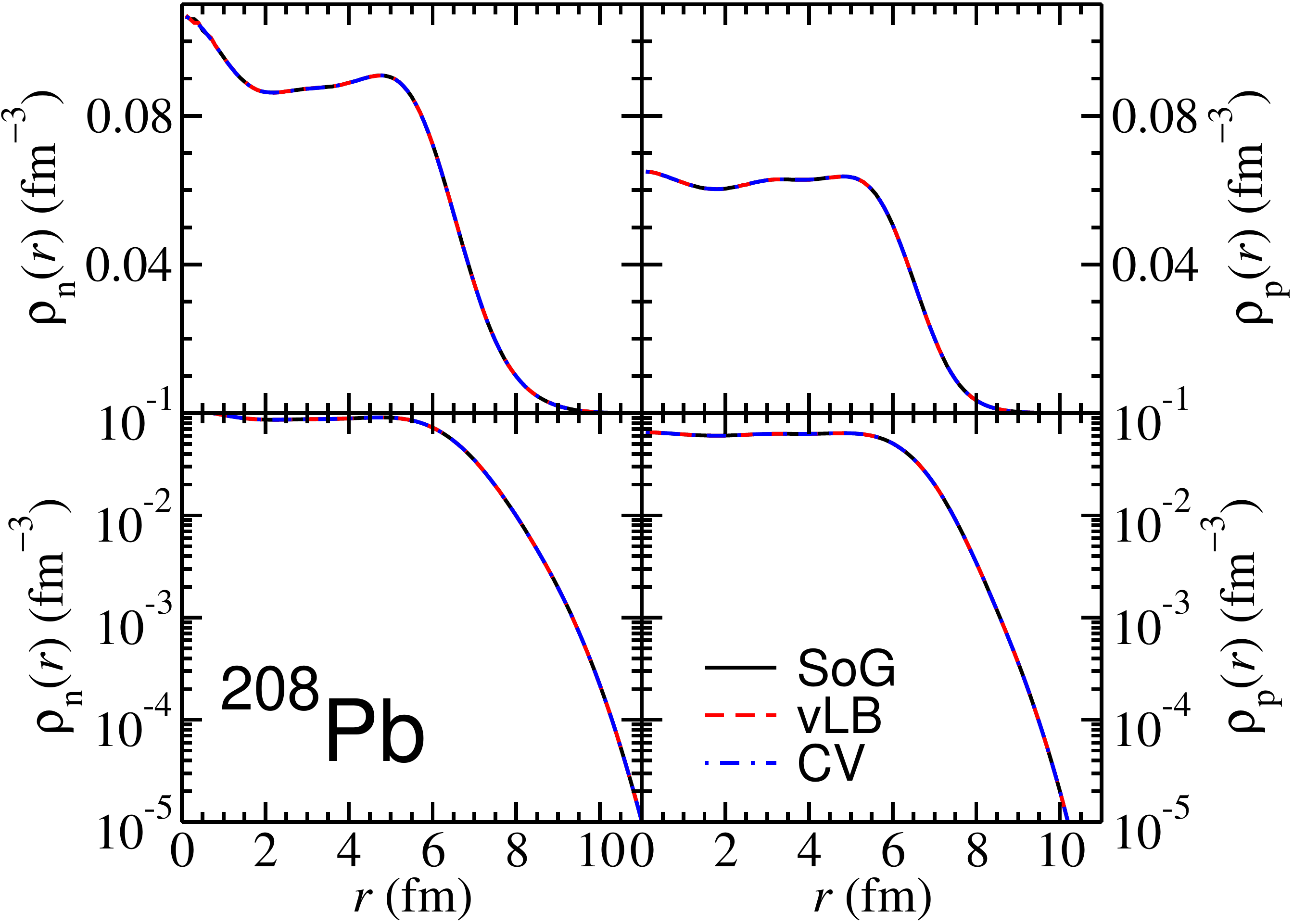}
\hfill  
\includegraphics[width=0.44\linewidth,clip=true]{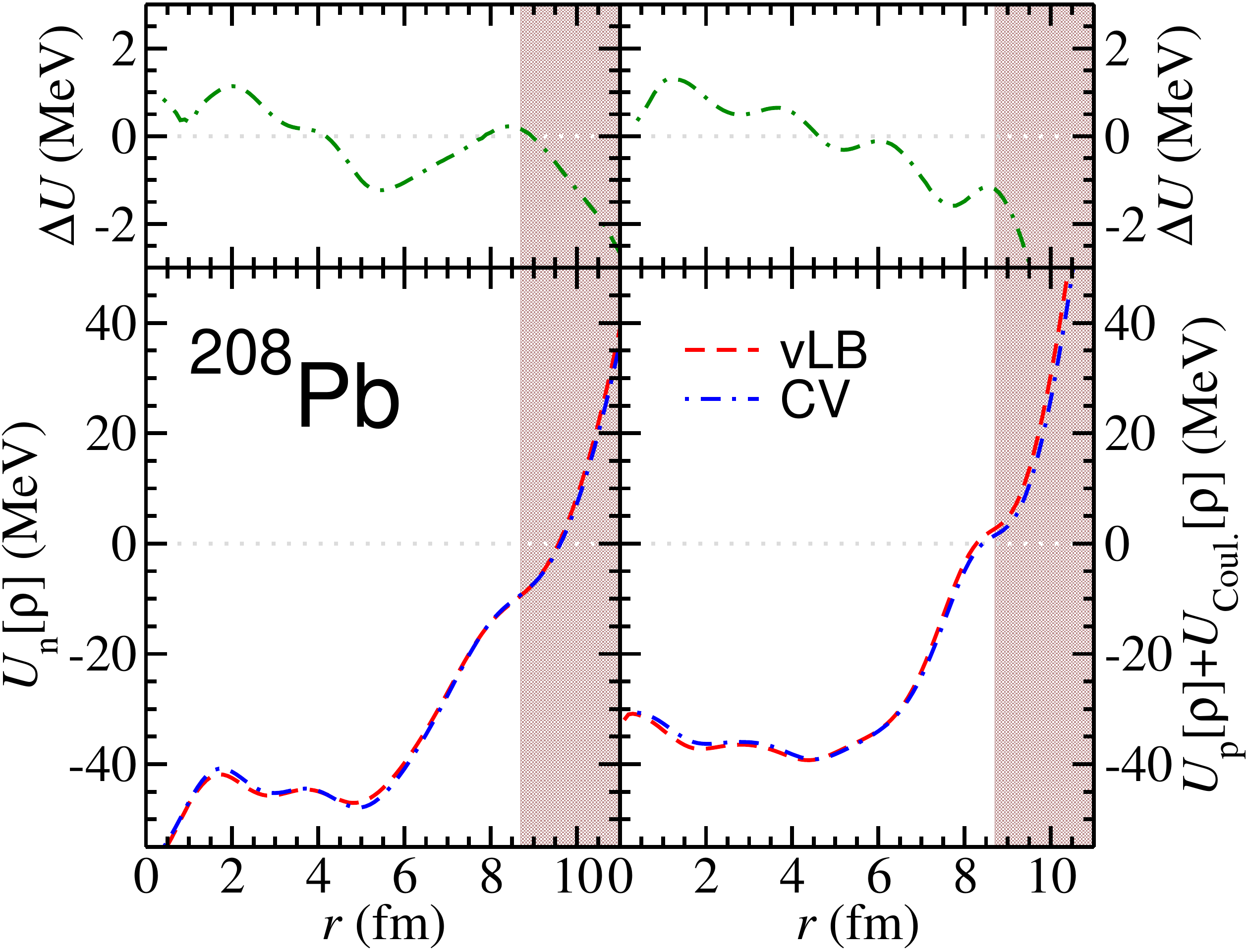}
\caption{(left figure) The neutron and proton densities (left panels) for the case of ${}^{208}$Pb  are displayed as a function of the radial coordinate, on a linear scale (top panels) and on a logarithmic scale (bottom panels). The target experimental densities \cite{DeVries}  labeled as SoG -- sum of Gaussians -- (black solid lines) are compared with those obtained  with the inversion methods vLB (red dashed lines) and CV (blue dot-dashed lines). (right figure)  The Kohn-Sham potentials  calculated for neutrons and protons with the inversion methods vLB (red dashed lines) and CV (blue dot-dashed lines)  are shown.  In the top panels, the  corresponding differences between the Kohn-Sham potentials $\Delta U=U_{\rm CV}-U_{\rm vLB}$ are displayed.}
\label{fig:208Pb_Den_Pot}
\end{figure*}

We have implemented the same procedure described in Sec. \ref{methods} in order to solve the IKS problem with
the input of the experimental densities. We have converged to KS densities that 
display a good agreement with the experimental densities. 
The agreement can be seen in the left panels of Fig.~\ref{fig:40Ca_Den_Pot} for ${}^{40}$Ca and Fig.~\ref{fig:208Pb_Den_Pot} for ${}^{208}$Pb. The relative differences 
found for the densities are of the same order of those found in the previous section. Since the differences between the vLB and CV densities and the target densities are not visible in detail from the figures, the maximum and the average of the absolute value of these differences are reported in Table \ref{table_diff_SOG}.

The Kohn-Sham potentials, shifted by using the experimental neutron and proton 
separation energies, obtained with the vLB and CV methods are also shown in Fig.~\ref{fig:40Ca_Den_Pot} for $^{40}$Ca and Fig.~\ref{fig:208Pb_Den_Pot} for $^{208}$Pb. The agreement between the two inversion methods is remarkable and of the same quality as that found in Sec. \ref{test} for the HF test cases (see the right panels in Figs.~\ref{fig:40Ca_Den_Pot} and \ref{fig:208Pb_Den_Pot}). However,  while the potentials in the inner part of the nuclei look very reasonable, they oscillate and then tend to increase without limit in the asymptotic region.
That is, the asymptotic behavior of the KS potentials at large distances is not the expected one. 
This has to be attributed to the Gaussian tail of the SoG density that both algorithms translate into a quadratic (i.e. harmonic oscillator-like) potential. To substantiate this interpretation, the regions corresponding to $r$ larger than the radius of the outermost  (second outermost) Gaussian in the case of $^{208}$Pb ($^{40}$Ca) are shown as shadowed areas in Figs.~\ref{fig:40Ca_Den_Pot} and \ref{fig:208Pb_Den_Pot}, respectively. The borders of these regions are clearly correlated with the change in slope of the potentials. As a consequence, our results for the experimentally derived KS potentials cannot be regarded as reliable in the tail of the potential.

We can conclude that the employed inversion procedure remains robust when experimental SoG densities are input and provides us with reliable information about the potential, except for its tail.

\begin{table}[hbt]
\centering
\begin{tabular}{r|cc|cc}
\hline
\hline  
Nucleus   & \multicolumn{2}{c|}{vLB}&\multicolumn{2}{c}{CV}\\
          &Max. & Aver. &Max. & Aver. \\
\hline
$^{40}$Ca~(p)  & 3.0 & 1.4& 0.9 & 0.2\\
$^{208}$Pb~(p) & 5.0 & 1.1& 0.2 & 0.04\\
$^{208}$Pb~(n) & 17.1 & 1.3& 0.4 & 0.1\\
\hline
\hline
\end{tabular}
\caption{Maximum (Max.) and average (Aver.) differences between the target experimental (SoG) and KS neutron (n) and proton (p) densities from the two inversion methods for the case of $^{40}$Ca and $^{208}$Pb. Numerical values are in $10^{-6}~\si{fm^{-3}}$.}
\label{table_diff_SOG}
\end{table}

\section{Conclusions and perspectives}
\label{conclusions}

We have addressed the inverse Kohn-Sham problem in the case of the atomic nucleus  for the first time, employing two well-known inversion methods  that have been used in other fields in physics \cite{JW:18}. The first method is based on an iterative procedure 
\cite{PhysRevA.49.2421} (vLB) and the second consists in a constrained minimization of the kinetic energy \cite{JW:18} (CV). 
We have applied  the inversion methods on the two  closed shell, spherical nuclei  ${}^{40}$Ca and ${}^{208}$Pb. 
We have first tested  the numerical algorithms deriving   the KS potential from a density obtained by a HF calculation with a Skyrme interaction
characterized by an effective mass close to the bare mass. We have verified that the resulting  KS potential is in good agreement  
with the HF potential. 
We have then applied the inversion methods to the experimentally derived densities of protons in ${}^{40}$Ca and ${}^{208}$Pb and neutrons in ${}^{208}$Pb. The consistency between vLB and CV remains remarkable, and the potentials obtained in the interior and at the surface of the nucleus 
appear to be reliable. 
On the other  hand, the parameterization of the experimental density  as a sum of Gaussians (SoG) used in this work, leads to difficulties 
in the tails of the potentials. The non-physical Gaussian tails probed by the algorithm at large distances 
translate into a harmonic oscillator-like potentials that diverge. 

\begin{itemize}
\item
Although attempting to use the outlined procedure for a larger set of nuclei, including
deformed ones, might be of interest, it is quite clear from the start that 
the mere experimental information about neutron and proton densities is
insufficient to deduce an effective KS potential. In fact, as we mentioned in the text above,
we know that a realistic nuclear EDF depends also on gradients of the densities, spin
densities and their combinations. These cannot be experimentally determined, and we
have to rely on {\em ab initio} calculations. As {\em ab initio} nuclear structure is
progressing, the first step to be undertaken should be to test the IKS when densities from
{\em ab initio} are input. This will allow to formulate the IKS scheme in a somehow
more general manner. As we mentioned in the text above, we envisage to proceed, at least, 
along two directions. We will try to fix the Equation of State (EoS) of uniform matter
directly from {\em ab initio} calculations, and formulate the IKS for finite nuclei in
such a way that only the gradient terms need to be extracted. At the same time, 
we will try to use the spin densities from {\em ab initio} to extract the spin part of
the effective potential. We are fully aware that in principle other kinds of densities should enter the game,
and further progress will eventually be needed; however, at this first stage, we will
stick to gradual steps and take for instance the spin-orbit and Coulomb parts of the EDF as
uncorrelated \cite{Baldo.08}. 
\item
A specific aspect, yet related to the previous point, is the issue of locality.
There is no guarantee that a purely local effective potential is the correct choice. 
In nuclear physics, local EDFs produce local potentials (like in the case of Skyrme EDFs) but
non-local EDFs and non-local potentials also exist (like in the case of Gogny); however, 
there is a way to re-parametrize non-locality in terms of a power expansion
in gradients as shown in Ref. \cite{PhysRevLett.105.122501}. Non-local densities from
{\em ab initio} calculations can be inspected to understand the degree of non-locality
which is needed.
\item
Last but most importantly, extracting an effective potential from IKS is not enough
to determine the quantity of real interest, that is the EDF itself. 
As shown by R. van Leeuwen and
E.J. Baerends in Ref. \cite{PhysRevA.51.170},
knowing the effective potential along a path of densities can give access to
the EDF. In few cases, we
can expect that {\em ab initio} calculations can explore system which are very close
to each other in terms of density distributions. One example are neutron drops
(see \cite{ZhaoPW2016,Shen2018b} and references therein), 
that are systems of neutrons confined by a harmonic
oscillator potential. We shall 
explore the possibility of other cases in which {\em ab initio} calculations 
can be performed for systems whose densities define a continuous path.
Last but not least,
checking whether this idea is related to the one introduced in Ref.~\cite{Dobaczewski_2016}, 
is also to be considered as a task to deal with.
\end{itemize}
Most likely, 
the most promising path to follow is probably the one in
which the IKS is used in parallel with other techniques to derive an EDF {\em ab initio}, 
as a way to fine-tune specific terms and not as a unique strategy. We envisage to
start soon to apply the IKS method to densities from {\em ab initio} approaches and
to understand how the current scheme can be generalized.

\section*{Acknowledgments}
Funding from the European Union's Horizon 2020 research and innovation programme under grant agreement No 654002 is acknowledged.

\bibliography{bibliography.bib}

\end{document}